\documentclass[a4paper,11pt]{article}
\usepackage{pos}
\newcommand{\lalat}{\ensuremath{\lambda_{\text{lat}}} }
\newcommand{\SO}[1]{\ensuremath{\text{SO(}#1\text{)}} }
\newcommand{\U}[1]{\ensuremath{\text{U(}#1\text{)}} }
\newcommand{\Uone}{\ensuremath{\text{U(1)}} }

\title{Large-$N$ limit of two-dimensional Yang--Mills theory with four supercharges}

\author*[a]{Navdeep Singh Dhindsa}
\author[b]{Raghav G.~Jha}
\author[a]{Anosh Joseph}
\author[c]{David Schaich}

\affiliation[a]{Department of Physical Sciences, Indian Institute of Science Education and Research - Mohali, \\
  Knowledge City, Sector 81, SAS Nagar, Punjab 140306, India}

\affiliation[b]{Perimeter Institute for Theoretical Physics, Waterloo, Ontario N2L 2Y5, Canada}

\affiliation[c]{Department of Mathematical Sciences, University of Liverpool, \\ Liverpool L69 7ZL, United Kingdom}

\emailAdd{navdeep.s.dhindsa@gmail.com}
\emailAdd{raghav.govind.jha@gmail.com}
\emailAdd{anoshjoseph@iisermohali.ac.in}
\emailAdd{david.schaich@liverpool.ac.uk}

\abstract{We study the two-dimensional Yang--Mills theory with four supercharges in the large-$N$ limit. By using thermal boundary conditions, we analyze the internal energy and the distribution of scalars. We compare their behavior to the maximally supersymmetric case with sixteen supercharges, which is known to admit a holographic interpretation. Our lattice results for the scalar distribution show no visible dependence on $N$ and the energy at strong coupling appears independent of temperature.}

\FullConference{%
 The 38th International Symposium on Lattice Field Theory, LATTICE2021
  26th-30th July, 2021
  Zoom/Gather@Massachusetts Institute of Technology
}

\begin{document}
\maketitle

\section{Introduction and motivation}
\label{Sec:Intro}

Large-$N$ studies of various supersymmetric Yang--Mills (SYM) theories can help us understand the non-perturbative nature of their holographic dual counterparts~\cite{Maldacena:1997re, Motl:1997th, Itzhaki:1998dd}. Regularizing these theories on a lattice would open up a first-principles approach to study these quantum field theories. However, it is not so straightforward to regularize these theories on a lattice (see Ref.~\cite{Schaich:2018mmv} for a recent review). Supersymmetric theories with a relatively larger number of supersymmetries can be discretized on a lattice with the help of twisting or orbifolding~\cite{Cohen:2003xe, Sugino:2004qd, Catterall:2004np}.

In this work we consider a two-dimensional SYM theory with four supercharges. A twisted version of this theory can be regularized on a lattice, with the help of a geometrical discretization scheme. The resultant lattice theory preserves one supercharge at finite lattice spacing~\cite{Catterall:2017xox}. The motivations for studying two-dimensional four-supercharge SYM theory on a lattice are the following. This theory is one of the simplest two-dimensional supersymmetric gauge theories to study on the lattice with a well-defined continuum limit. It does not exhibit the numerical sign problem when the continuum limit is properly taken~\cite{Catterall:2011aa, Galvez:2011ief, Mehta:2011ud, Hanada:2010qg, Catterall:2017xox}. The maximally supersymmetric counterpart of this theory, i.e., the two-dimensional theory with sixteen supercharges, has a holographic dual. Although the theory under consideration does not have a gravity dual, we study this to understand whether it has features that belong to the same universality class as that of its sixteen-supercharge counterpart. 

Our goal, in addition to studying the thermodynamics of the theory, is to look for the existence of a bound state, at large $N$, in which the scalar fields clump around the origin, in spite of the existence of a classical flat direction. This bound state is metastable at finite $N$ and becomes stable at large $N$. It is similar to the one found in the sixteen supercharge model \cite{Anagnostopoulos:2007fw}. There, through the gauge-gravity duality, the scalar fields represent the collective coordinates of the D-branes and thus the bunch of the D-branes is nothing but the bound state of scalars. From a purely field theoretical point of view the very existence of this bound state is nontrivial and this is also the reason why we resort to numerical computation. For computing the observables, we impose anti-periodic boundary conditions for fermions along the temporal direction. It has been previously observed that with periodic boundary conditions the scalars clump around origin even in the presence of flat directions~\cite{Hanada:2009hq}.

We present preliminary results on the energy density and distribution of scalars for this theory. These results are obtained for a range of temperatures with several numbers of colors $N \leq 12$. In Sec.~\ref{Sec:Theory} we briefly discuss two-dimensional $\mathcal{N} = (2, 2)$ SYM and its lattice construction. The results are discussed in Sec.~\ref{Sec:Results}. Future directions are discussed in Sec.~\ref{Sec:Future_directions}.

\section{Two-dimensional $\mathcal{N} = (2, 2)$ SYM}
\label{Sec:Theory}

To study our target theory in a flat Euclidean spacetime, we will use the method of topological twisting. The two-dimensional $\mathcal{N} = (2, 2)$ SYM is constructed from the dimensional reduction of $\mathcal{N} = 1$ SYM in four dimensions. The parent theory has a global symmetry group $G = \SO{4}_E \times \Uone$, where $\SO{4}_E$ is the Euclidean rotation symmetry and \Uone is the chiral symmetry. After dimensionally reducing this to two dimensions, the global symmetry group becomes $G = \SO{2}_E \times \SO{2}_{R_1}\times \Uone_{R_2}$, where $\SO{2}_E$ is the Euclidean rotation symmetry, $\SO{2}_{R_1}$ is the rotation symmetry along the reduced dimensions and $\Uone_{R_2}$ is the chiral symmetry of the theory. To maximally twist the theory in $d$ dimensions,\footnote{See Ref.~\cite{Joseph:2011xy} for a review of maximally twisted SYM theories on a lattice.} the $R$ symmetry group should contain $\SO{d}$ as a subgroup. The $R$ symmetry group in our target theory has two $\SO{2}$ factors as \Uone is also locally equivalent to $\SO{2}$. Hence we can twist this theory by combining $\SO{2}_E$ with either $\Uone_{R_2}$ or $\SO{2}_{R_1}$, which are called the $A$-model and $B$-model twists, respectively. We work with the $B$-model twist for which the twisted rotation group takes the form
\begin{equation}
  \SO{2}' =  \text{diag} \Big(\SO{2}_E \times \SO{2}_{R_1}\Big).
\end{equation}

The theory has four bosonic and four fermionic degrees of freedom. It also has four real supercharges. After twisting, the fermions and supercharges decompose into integer-spin representations of $\SO{2}'$. In the twisted formalism, the gauge field ($A_a$) and scalars ($X_a$) transform identically, and hence we can combine them to form a complexified gauge field $\mathcal{A}_a = A_a + i X_a$. Thus the action in the twisted form is composed of a complexified gauge field $\mathcal{A}_a$ and the twisted fermions $\eta$, $\psi_a$, and $\chi_{ab} = -\chi_{ba}$. The action is given by
\begin{equation}
  S = \frac{N}{4\lambda} \mathcal{Q} \int d^2x ~\text{Tr} \left(\chi_{ab} \mathcal{F}_{ab} + \eta \left[ \overline{\mathcal{D}}_a, \mathcal{D}_a \right] - \frac{1}{2} \eta d \right),
\end{equation}
where $\lambda$ is the 't~Hooft coupling. The complexified field strengths are $\mathcal{F}_{ab} = \left[\mathcal{D}_a, \mathcal{D}_b\right]$ and $\overline{\mathcal{F}}_{ab} = \left[\overline{\mathcal{D}}_a, \overline{\mathcal{D}}_b\right]$, with $\mathcal{D}_a = \partial_a + \mathcal{A}_a$ and $\overline{\mathcal{D}}_a = \partial_a + \overline{\mathcal{A}}_a$ denoting the complexified covariant derivatives. The nilpotent supercharge $\mathcal{Q}$ acts on the twisted fields in the following way:
\begin{align}
  \mathcal{QA}_a & = \psi_a &
  \mathcal{Q\overline{\mathcal{A}}}_a & = 0 &
  \mathcal{Q\psi}_a & = 0 \nonumber \\
  \mathcal{Q\chi}_{ab} & = -\overline{\mathcal{F}}_{ab} &
  \mathcal{Q} \eta & = d &
  \mathcal{Q}d & = 0,
\end{align}
where $d$ is a bosonic auxiliary field with equation of motion $d = \left[\overline{\mathcal{D}}_a, \mathcal{D}_a\right]$.
After performing the $\mathcal{Q}$ variation on the action and integrating out the auxiliary field $d$, we get
\begin{equation}
  S = \frac{N}{4\lambda} \int d^2x~\text{Tr} \left(-\overline{\mathcal{F}}_{ab}\mathcal{F}_{ab} + \frac{1}{2} \left[ \overline{\mathcal{D}}_a, \mathcal{D}_a \right]^2 - \chi_{ab}\mathcal{D}_{[a~} \psi_{~ b]} - \eta \overline{\mathcal{D}}_a \psi_a \right).
\end{equation}

Now, upon using the geometrical discretization scheme, we arrive at the arrangement in which the two-dimensional $\mathcal{N} = (2, 2)$ SYM lives on a two-dimensional square lattice spanned by two orthogonal unit vectors. The complexified gauge field is mapped to a complexified Wilson link, $\mathcal{A}_a(x) \rightarrow \mathcal{U}_a(n)$ on the lattice. Its fermionic superpartner $\psi_a(n)$ also transforms as a link variable. The field $\eta(n)$ lives on a site. The 2-form field $\chi_{ab}$ lives on the diagonal of the unit cell. The orientations of the fields on the lattice ensure gauge invariance.

The continuum derivatives are replaced by covariant difference operators according to the geometrical discretization scheme, prescribed in Ref.~\cite{Damgaard:2008pa}. The curl-like covariant derivatives and divergence-like covariant derivatives become forward- and backward-difference operators, respectively. We have
\begin{align}
  \overline{\mathcal{D}}_a^{(-)} f_a(n) & = f_a(n) \overline{\mathcal{U}}_a(n) - \overline{\mathcal{U}}_a(n - \hat{\mu}_a) f_a(n - \hat{\mu}_a), \nonumber \\
  \mathcal{D}_a^{(+)} f_b(n) & = \mathcal{U}_a(n) f_b(n + \hat{\mu}_a) - f_b(n) \mathcal{U}_a(n + \hat{\mu}_b).
\end{align}

The lattice action then becomes
\begin{eqnarray}
  S &=& \frac{N}{4\lalat} \sum_n \text{Tr} \Bigg[-\overline{\mathcal{F}}_{ab}(n) \mathcal{F}_{ab}(n) + \frac{1}{2} \left(\overline{\mathcal{D}}_a^{(-)} \mathcal{U}_a(n)\right)^2 \nonumber \\
  && - \chi_{ab}(n) \mathcal{D}^{(+)}_{~[a}\psi^{\ }_{~b]}(n)-\eta(n) \overline{\mathcal{D}}_a^{(-)}\psi_a(n)\Bigg].
\end{eqnarray}
In addition, to control the flat directions in the theory, we add a scalar potential term to the action with a tunable parameter $\mu$. Hence the complete action is given by
\begin{align}
  S_{\text{total}} = S + \frac{N}{4\lalat} \mu^2 \sum_{n, a} \text{Tr} \left(\overline{\mathcal{U}}_a(n) \mathcal{U}_a(n) - \mathbb{I}_N\right)^2.
\end{align}

\section{Preliminary results}
\label{Sec:Results}

The theory is analyzed on symmetric lattices with $a N_t = \beta = L = a N_x$, where $\beta$ and $L$ are dimensionful temporal and spatial extents, respectively. The lattice spacing is denoted by $a$, and $N_t$ and $N_x$ are the number of lattice sites along the temporal and spatial directions, respectively. Anti-periodic boundary conditions are imposed along the temporal direction for fermions. The fields are made dimensionless in terms of the 't~Hooft coupling $\lambda$. Hence the dimensionless temporal (and spatial) extent becomes $r_t = \sqrt{\lambda}\beta = 1 / t$, which also serves as an effective coupling and inverse dimensionless temperature. We set the scalar potential parameter to be $\mu = \zeta \frac{r_t}{N_t} = \zeta \sqrt{\lambda} a = \zeta \sqrt{\lalat}$, making $\zeta$ the parameter we tune.
The flat directions of the continuum theory are recovered in the limit $\zeta \to 0$.

The calculations are performed using the publicly available software\footnote{\href{https://github.com/daschaich/susy}{github.com/daschaich/susy}} described in Ref.~\cite{Schaich:2014pda}. For the preliminary analysis, we consider several numbers of colors up to $N = 12$ and lattice sizes up to $32^2$ over a range of temperatures.

\begin{figure}[htp]
  \begin{center}
    \includegraphics[width=.48\textwidth]{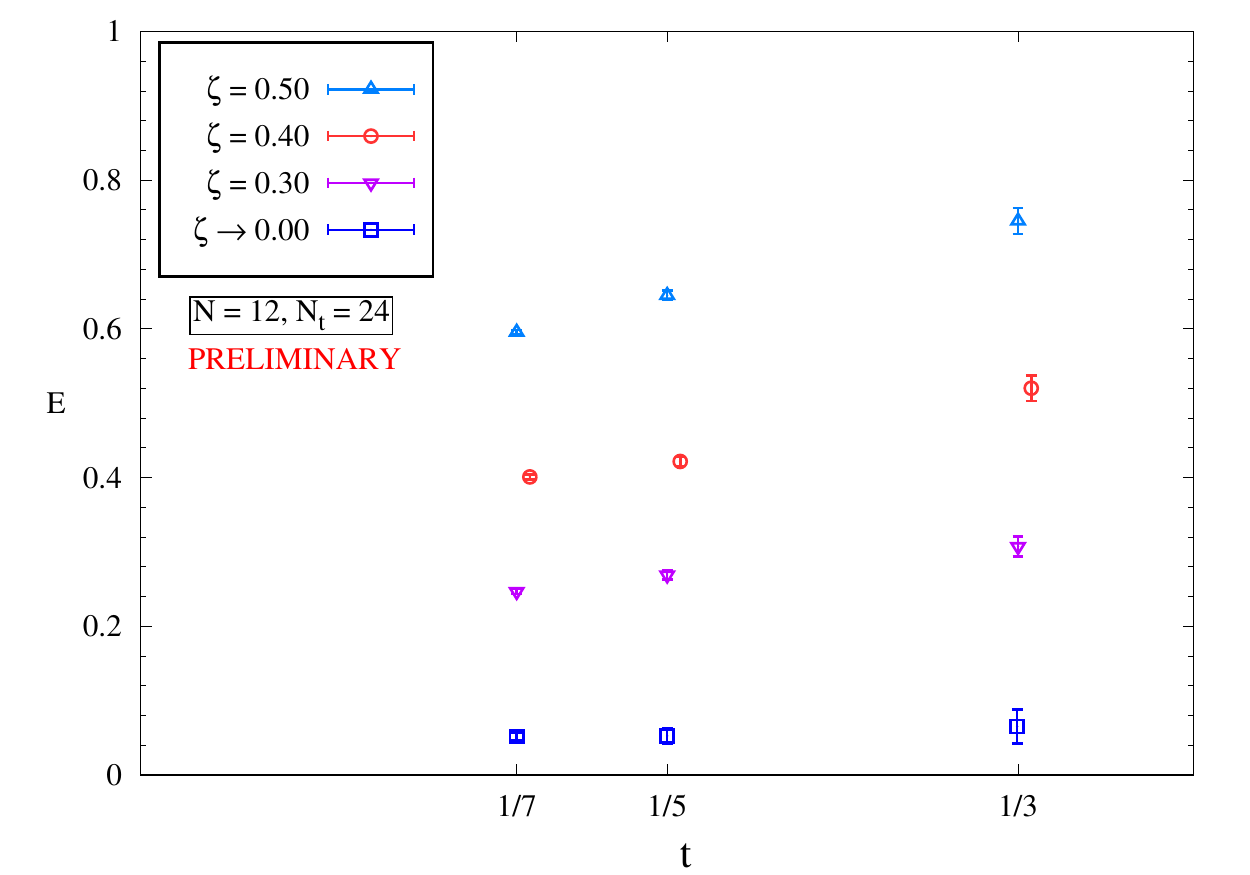}
    \includegraphics[width=.48\textwidth]{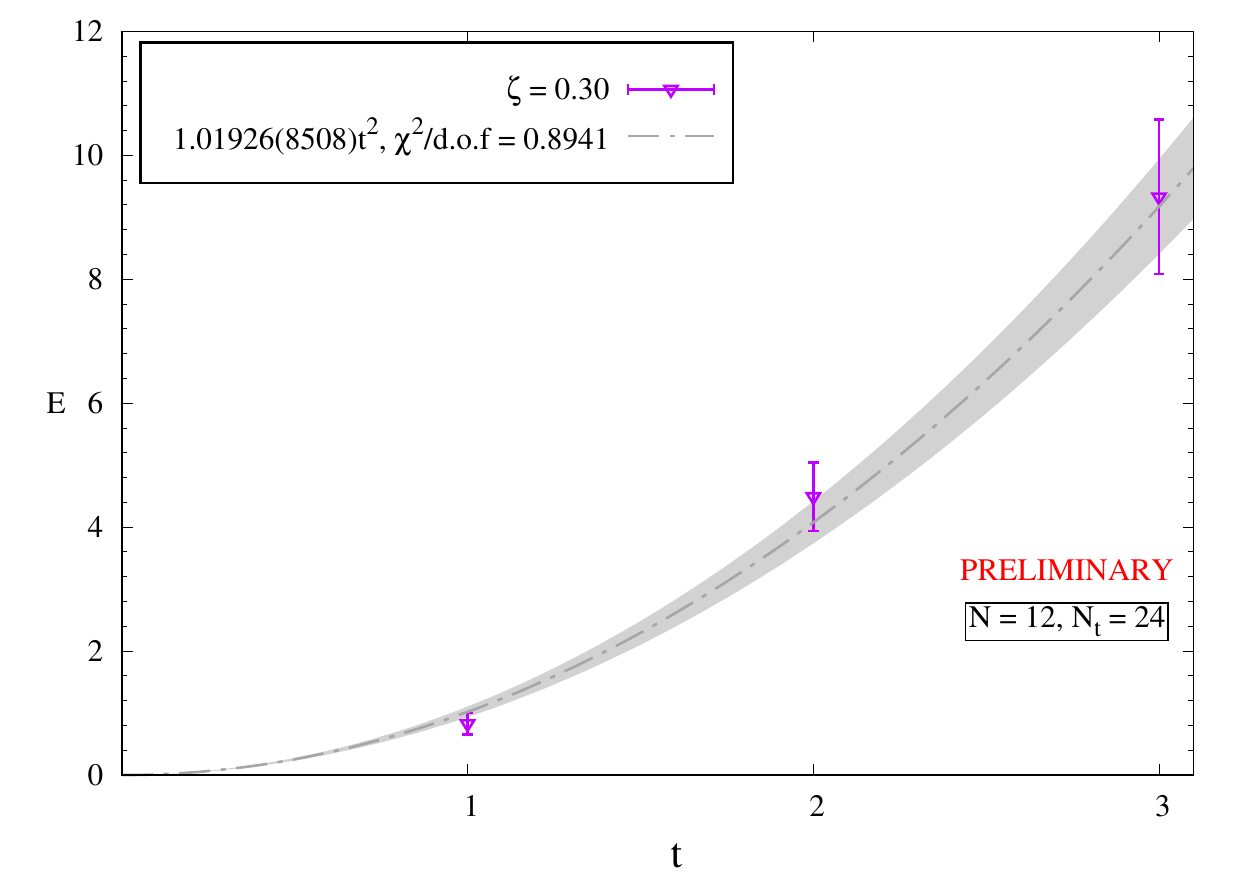}
  \end{center}
  \caption{Energy density (Eq.~\ref{eq:energy}) for $t < 1$ with different $\zeta$ values (left) and $t \ge 1$ with $\zeta = 0.3$ (right). Both plots are for $N = 12$ and lattice size $24^2$.}
  \label{Fig:barS}
\end{figure}

We first report on the energy density
\begin{equation}
  \label{eq:energy}
  E = \frac{3}{\lalat}\left(1-\frac{2}{3N^2}S_B\right),
\end{equation}
where $S_B$ is the bosonic action averaged over the lattice volume and \lalat is the dimensionless 't~Hooft coupling.
The results for temperatures $t < 1$ and $t \ge 1$ are shown in Fig.~\ref{Fig:barS}. At high temperatures (weak effective couplings $r_t$), the $E \propto t^2$ behavior of the energy density is the same as that of the maximally supersymmetric theory as reported in Refs.~\cite{Catterall:2017lub, Jha:2017zad}, with a different coefficient. However, at low temperatures (strong effective couplings $r_t$) the $\zeta \to 0$ energy density appears to be independent of the temperature, in contrast to the maximally supersymmetric case where a cubic temperature dependence is known~\cite{Wiseman:2013cda}.

\begin{figure}[htp]
  \begin{center}
    \includegraphics[width=.48\textwidth]{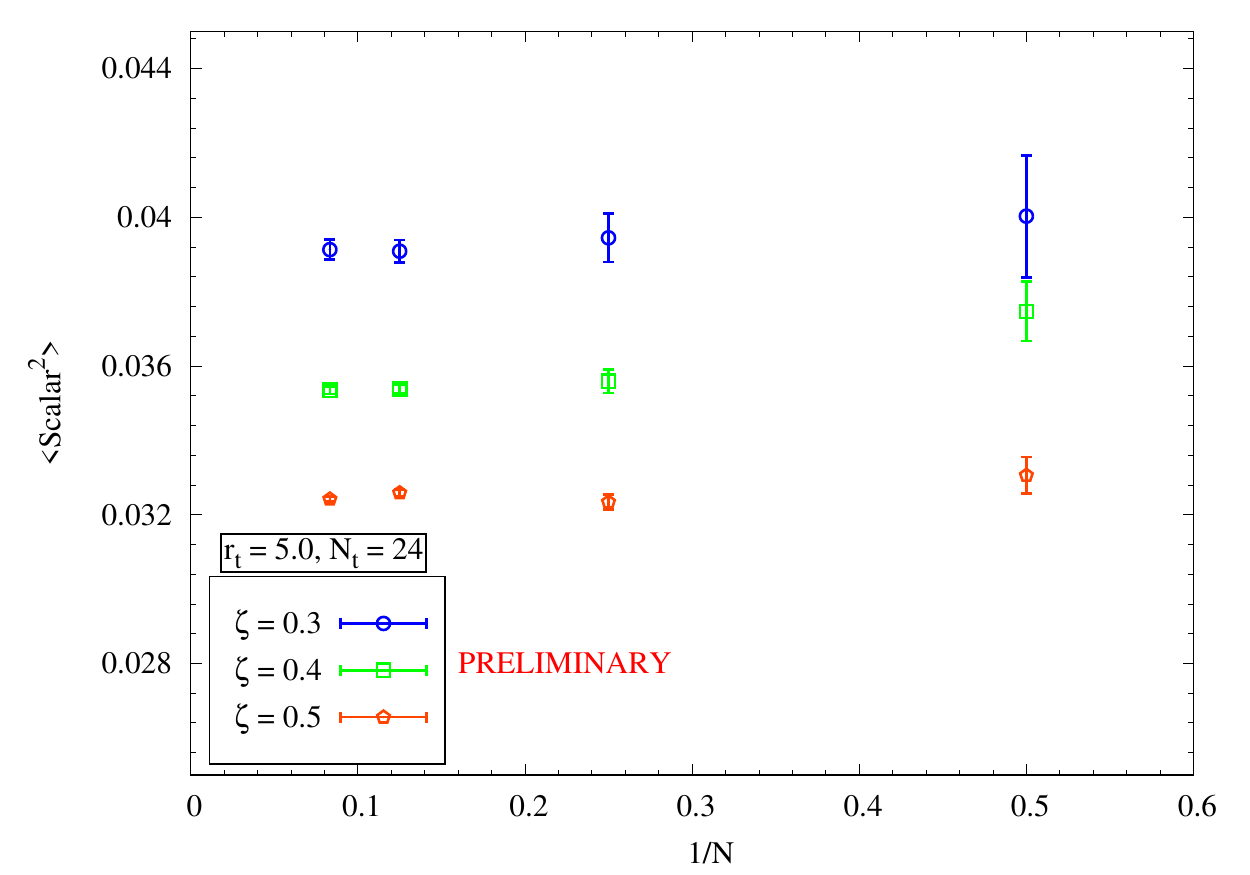}
    \includegraphics[width=.48\textwidth]{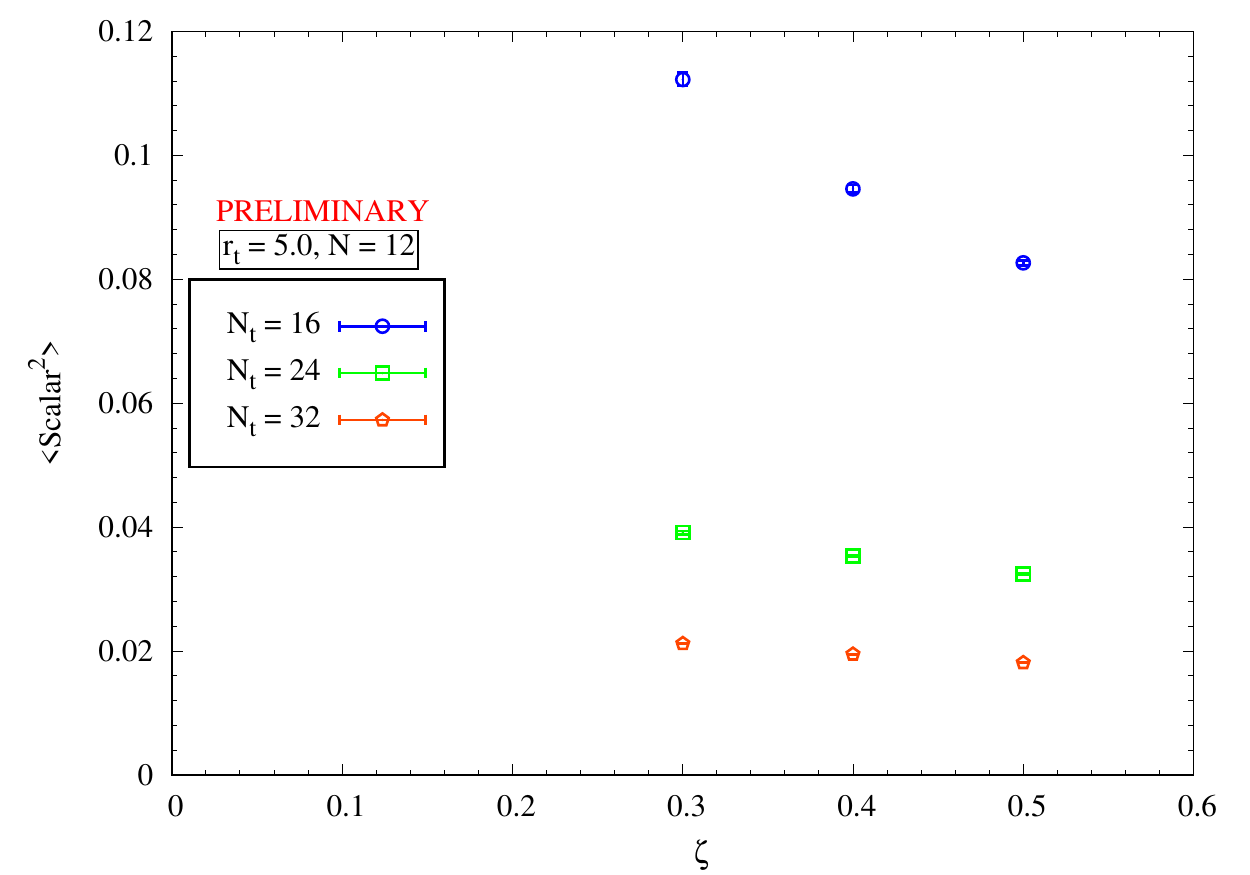}
  \end{center}
  \caption{The dependence of the scalar extent (Eq.~\ref{eqn:scalar}) on $N \in \left\{2, 4, 8, 12\right\}$ for different $\zeta$ at fixed $r_t = 5$ on $24^2$ lattices (left), and its dependence on $\zeta$ for different lattice sizes at fixed $r_t = 5$ and $N = 12$ (right).}
  \label{Fig:Tr_X2}
\end{figure}

The other observable we study is the extent of the scalars. It captures the existence of bound state in the theory. It is defined as
\begin{equation}
  \left\langle \text{Scalar}^2 \right\rangle = \left\langle \frac{1}{N N_t N_x}~\sum_{i=1}^{N_t N_x}~\text{Tr}(X_i^2) \right\rangle.
  \label{eqn:scalar}
\end{equation}
Our preliminary analyses of $\left\langle \text{Scalar}^2 \right\rangle$ at $r_t = 5$ for different $N$ and $N_t$ values are shown in Fig.~\ref{Fig:Tr_X2}.

Considering the large-$N$ limit in the left-panel plot of Fig.~\ref{Fig:Tr_X2}, we observe no visible dependence on $N$, for a relatively larger lattice with $N_t = 24$.
These results suggest that the scalars seem to clump around the origin even for small $\U{N}$ gauge groups, provided we are working with sufficiently large lattices. For smaller lattices, say $N_t = 16$, the right-panel plot in Fig.~\ref{Fig:Tr_X2} suggests that the scalars may not converge to a definite value in the $\zeta \to 0$ limit.

\begin{figure}[htp]
  \begin{center}
    \includegraphics[width=.7\textwidth]{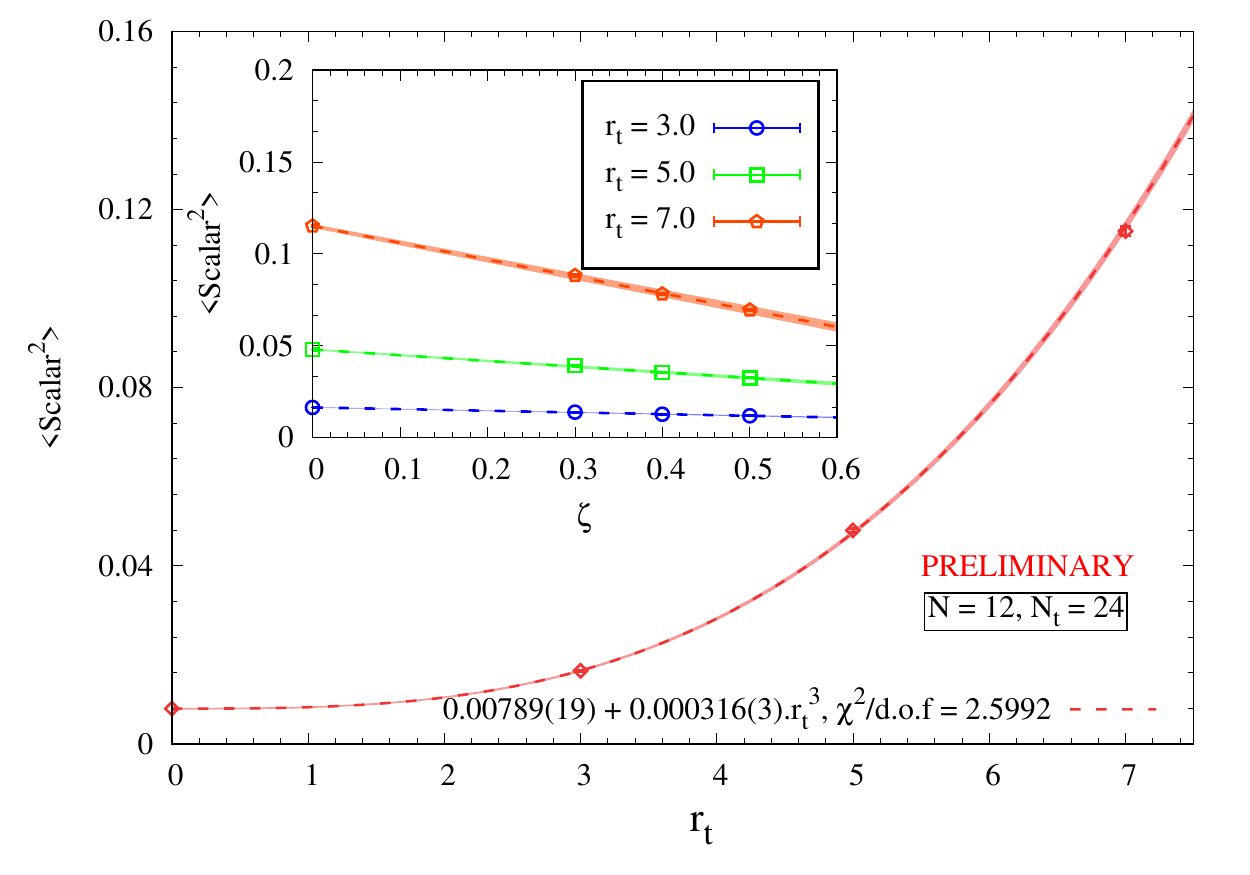}
  \end{center}
  \caption{The observable $\left\langle \text{Scalar}^2 \right\rangle$ given by Eq.~\eqref{eqn:scalar} is shown in the $\zeta \to 0$ limit for various couplings. In the inset, we show how the $\zeta \to 0$ limit is taken. The results are for $N = 12$ on $24^2$ lattices.}
  \label{Fig:Tr_X2_Final}
\end{figure}

For $N = 12$ and $N_t = 24$ the behavior of the scalars in the low-temperature region of strong effective couplings $r_t$ is shown in Fig.~\ref{Fig:Tr_X2_Final}.
The distribution of scalars for a fixed $r_t$ is estimated by considering several $\zeta$ values and taking the $\zeta \to 0$ limit.
As the $r_t \to 0$ limit is taken $\left\langle \text{Scalar}^2 \right\rangle$ approaches a nonzero value, which indicates that the scalars do not show any run-away behavior in the presence of flat directions and hence the existence of bound state at finite temperature is observed. The $\sim\! r_t^3$ dependence we observe is very different from the $\sim\! 1/r_t$ behavior predicted for the maximally supersymmetric theory.

\section{Conclusions and way forward}
\label{Sec:Future_directions}

The preliminary analysis reported in this work indicates the existence of the bound states of scalars at finite temperatures for two-dimensional $\mathcal{N} = (2, 2)$ SYM. Even though we need large lattices to construct the bound state (and to control the sign problem~\cite{Catterall:2011aa, Hanada:2010qg, Catterall:2017xox}), we observe that there is no significant dependence on the gauge group $\U{N}$. Even modest values of $N \gtrsim 4$ appear to provide access to the large-$N$ limit of the theory given that we have data on large enough lattices.

By looking at the temperature dependence of the energy density and the distribution of the scalars we can say that the two-dimensional $\mathcal{N} = (2, 2)$ SYM theory seems to behave very differently at low temperatures compared to its maximal $\mathcal N = (8, 8)$ counterpart.

We also want to investigate whether there is a deconfinement transition in the $\mathcal N = (2, 2)$ theory as we move from small to large lattice sizes. In the sixteen-supercharge theory, there is a phase transition which is dual to a transition between different black hole solutions~\cite{Catterall:2017lub, Jha:2017zad}. We have seen signs of the $\mathcal N = (2, 2)$ SYM center symmetry being broken for small lattices (high temperatures) and restored for large lattices, but more careful analysis is needed. We are continuing our investigations by considering larger lattice sizes to do systematic studies of continuum extrapolations, as well as obtaining more robust results for the dependence on the number of colors over a wider range of temperatures.

\acknowledgments

NSD thanks the Council of Scientific and Industrial Research (CSIR), Government of India, for the financial support through a research fellowship (Award No.~{09/947(0119)/2019-EMR-I}).
RGJ would like to thank Masanori Hanada for discussions. Research at Perimeter Institute is supported in part by the Government of Canada through the Department of Innovation, Science and Economic Development and by the Province of Ontario through the Ministry of Colleges and Universities.
The work of AJ was supported in part by the Start-up Research Grant (No.~{SRG/2019/002035}) from the Science and Engineering Research Board (SERB), Government of India, and in part by a Seed Grant from the Indian Institute of Science Education and Research (IISER) Mohali.
DS was supported by UK Research and Innovation Future Leader Fellowship {MR/S015418/1} and STFC grant {ST/T000988/1}.
Numerical calculations were carried out at the University of Liverpool.

\bibliographystyle{JHEP}
\bibliography{lattice21}

\end{document}